\documentclass[12pt]{iopart}

\usepackage{graphicx}
\usepackage{overpic}
\usepackage{iopams}  

\begin{document}

\title[Ground-state statistics from annealing algorithms]{Ground-state
statistics from annealing algorithms: Quantum vs classical approaches}

\author{Yoshiki Matsuda}
\address{Department of Physics, Tokyo Institute of Technology,
Oh-okayama, Meguro-ku, Tokyo 152-8551, Japan}

\author{Hidetoshi Nishimori}
\address{Department of Physics, Tokyo Institute of Technology,
Oh-okayama, Meguro-ku, Tokyo 152-8551, Japan}

\author{Helmut G. Katzgraber}
\address{Theoretische Physik, ETH Zurich, CH-8093 Zurich,
         Switzerland}
\address{Department of Physics, Texas A\&M University, College Station,
         Texas 77843-4242, USA}

\begin{abstract}

We study the performance of quantum annealing for systems with
ground-state degeneracy by directly solving the Schr\"odinger equation
for small systems and quantum Monte Carlo simulations for larger
systems.  The results indicate that naive quantum annealing using
a transverse field may not be well suited to identify all degenerate
ground-state configurations, although the value of the ground-state
energy is often efficiently estimated.  An introduction of quantum
transitions to all states with equal weights is shown to greatly
improve the situation but with a sacrifice in the annealing time.
We also clarify the relation between the spin configurations in the
degenerate ground states and the probabilities that those states are
obtained by quantum annealing. The strengths and weaknesses of quantum
annealing for problems with degenerate ground states are discussed
in comparison with classical simulated annealing.

\end{abstract}


\maketitle

\section{Introduction}
\label{sec:intro}

Quantum annealing (QA) \cite{kadowaki:98,finnila:94} is the
quantum-mechanical version of the simulated annealing (SA) algorithm
\cite{kirkpatrick:83} to study optimization problems.  While the
latter uses the slow annealing of (classical) thermal fluctuations
to obtain the ground state of a target Hamiltonian (cost function),
the former uses quantum fluctuations. An extensive body of numerical
\cite{das:05,santoro:02,martonak:02,santoro:06,das:08,young:08} as
well as analytical \cite{morita:08} studies show that QA is generally
more efficient than SA for the ground-state search (optimization)
of classical Hamiltonians of the Ising type.  This fact does not
immediately imply, however, that SA will soon be replaced by QA in
practical applications because the full implementation of QA needs
an efficient method for solving the Schr\"odinger equation for large
systems, a task optimally achievable only on quantum computers. Given
continuing progress in the implementability of quantum computers,
we thus continue to study the theoretical efficiency and the limit
of applicability of QA using small-size prototypes and classical
simulations of quantum systems, following the general spirit of quantum
information theory.  So far, almost all problems studied with QA have
been for nondegenerate cases, and researchers have not paid particular
attention to the role played by degeneracy.  This question, however,
needs careful scrutiny because many practical problems have degenerate
ground states.  The present paper is a partial report on these efforts
with a focus on the efficiency of QA when the ground state of the
studied model is degenerate, i.e., when different configurations of
the degrees of freedom yield the same lowest-possible energy.

If the goal of the minimization of a Hamiltonian (cost function) of a
given problem is to obtain the ground-state energy (minimum of the cost
function), it suffices to reach one of the degenerate ground states,
which might often be easier than an equivalent nondegenerate problem
because there are many states that are energetically equivalent.
If, on the other hand, we are asked to identify all (or many of)
the degenerate ground-state configurations (arguments of the cost
function which minimize it) and not just the lowest value of the
energy, we have to carefully check if all ground states can be found.
This would thus mean that the chosen algorithm can reach all possible
ground-state configurations ergodically. Such a situation would happen
if, for instance, we want to compute the ground-state entropy or we
may need the detailed spin configuration of a spin-glass system to
understand the relationship between the distribution of frustration
and the ground-state configurations.

We have investigated this problem for a few typical systems with
degeneracy caused by frustration effects in the interactions between
Ising spins. Our results indicate that naive QA using transverse
fields is not necessarily well suited for the identification of all
the degenerate ground states, i.e., the method fails to find certain
ground-state configurations independent of the annealing rate.  This is
in contrast to SA, with which all the degenerate states are reached
with almost equal probability if the annealing rate of the temperature
is sufficiently slow. Nevertheless, when only the ground-state energy
is needed, QA is found to be superior to SA for some example systems.

The present paper is organized as follows: Section
\ref{sec:schroedinger} describes the solution of a small system by
direct diagonalization and numerical integration of the Schr\"odinger
equation.  Section \ref{sec:large} is devoted to the studies of larger
degenerate systems via quantum Monte Carlo simulations, followed by
concluding remarks in section \ref{sec:concl}.

\section{Schr\"odinger dynamics for a small system}
\label{sec:schroedinger}

\begin{figure}[h]
\scalebox{0.35}{\input{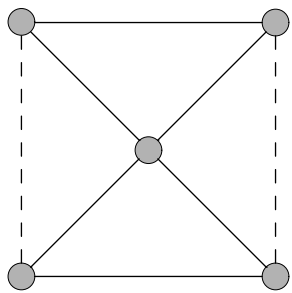}}
\begin{minipage}[b]{28pc}\caption{\label{fig:five-spins}
Five-spin toy model studied. Full lines denote ferromagnetic
interactions ($J_{ij}=1$) while dashed lines stand for
antiferromagnetic interactions ($J_{ij}=-1$). Because of the
geometry of the problem the system has a degenerate ground state
by construction.
}
\end{minipage}
\end{figure}

It is instructive to first study a small-size system by a direct
solution of the Schr\"odinger equation, both in stationary and
nonstationary contexts.  The classical optimization problem for this
purpose is chosen to be a five-spin system with interactions as shown
in figure \ref{fig:five-spins}.

\begin{figure}
\begin{center}
\input{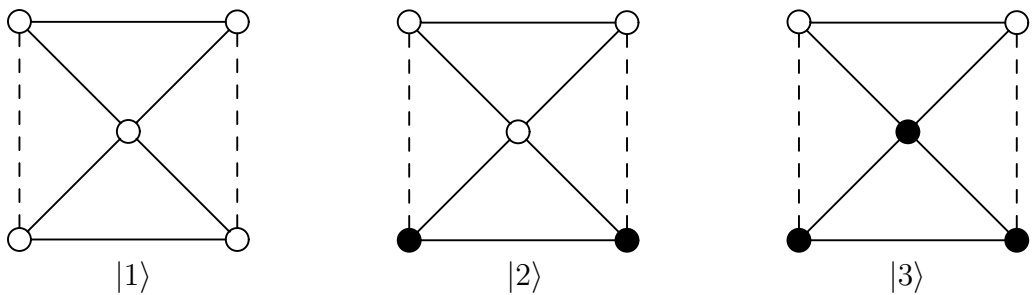}
\end{center}
\caption{
Nontrivial degenerate ground states of the toy model shown in figure
\ref{fig:five-spins}. Filled and open circles denote up and down
spins, respectively.  The other three ground states $|\bar{1}\rangle$,
$|\bar{2}\rangle$, and $|\bar{3}\rangle$ are obtained from $|1\rangle$,
$|2\rangle$, and $|3\rangle$ by reversing all spins.
\label{fig:six-ground-states}
}
\end{figure}

The Hamiltonian of this system is given by
\begin{equation}
  H_0=-\sum_{\langle ij\rangle}J_{ij} \sigma_i^z \sigma_j^z ,
  \label{H0}
\end{equation}
where the sum is over all nearest-neighbour interactions $J_{ij} = \pm
1$ and $\sigma_i^z$ denote Ising spins parallel to the $z$-axis. The
system has six degenerate ground states, three of which are shown in
figure \ref{fig:six-ground-states}.
We apply a transverse field
\begin{equation}
H_1=-\sum_{i} \sigma_i^x
\label{H1}
\end{equation}
to the system $H_0$ to induce a quantum transition between classical
states. The total Hamiltonian $H(t)$ changes from $H_1$ at $t=0$
to $H_0$ at $t=\tau$, i.e., 
\begin{equation}
H(t)=\left( 1-\frac{t}{\tau}\right) H_1+\frac{t}{\tau}H_0 .
\label{full-Hamiltonian}
\end{equation}
For large $\tau$ the system is more likely to follow the instantaneous
ground state according to the adiabatic theorem. If the target
optimization Hamiltonian $H_0$ had no degeneracy in the ground state,
the simple adiabatic evolution ($\tau\gg 1$) would drive the system
from the trivial initial ground state of $H_1$ to the nontrivial
final ground state of $H_0$ (solution of the optimization problem).

\begin{figure}[h]
\includegraphics[width=.45\linewidth]{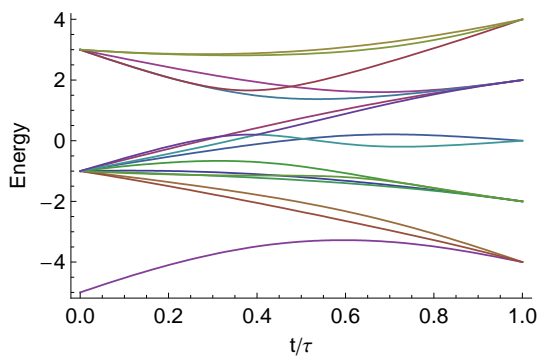}\hspace{2pc}
\begin{minipage}[b]{18pc}\caption{\label{fig:e-spectrum}
Instantaneous energy spectrum of the five-spin system depicted in
figure \ref{fig:five-spins}. For simplicity we have omitted the energy
levels that are not reachable from the ground state due to different
symmetry properties.
}
\end{minipage}
\end{figure}

\begin{figure}[htb]
\begin{minipage}{0.50\hsize}
\begin{center}
\includegraphics[width=1\linewidth]{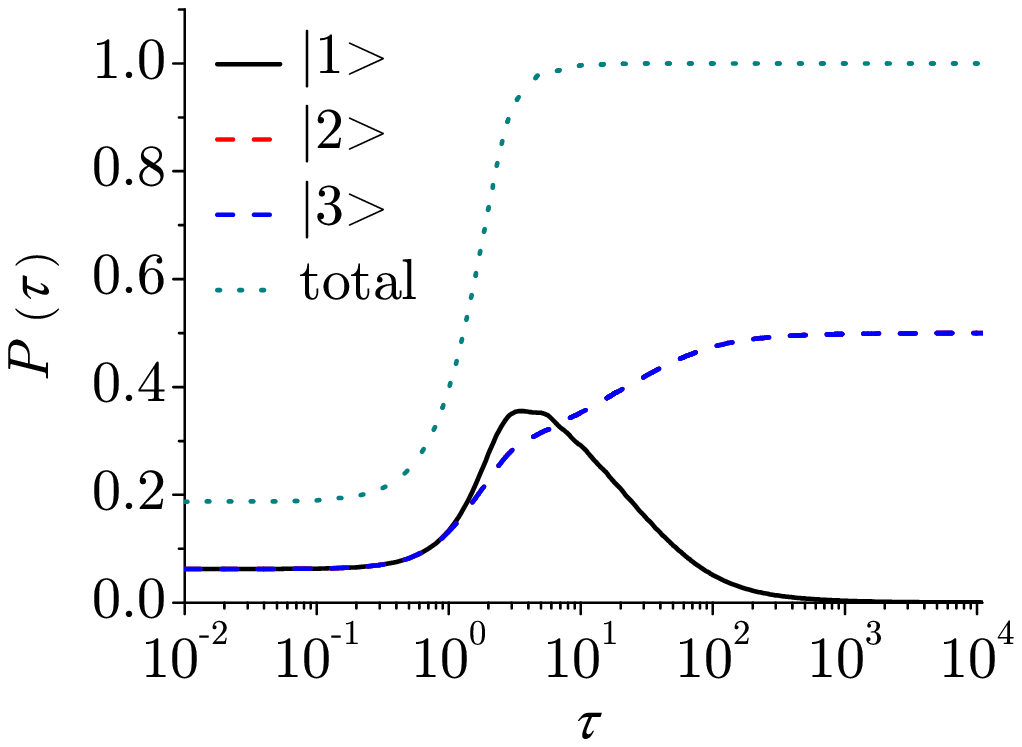}
\end{center}
\end{minipage}
\begin{minipage}{0.50\hsize}
\begin{center}
\includegraphics[width=1\linewidth]{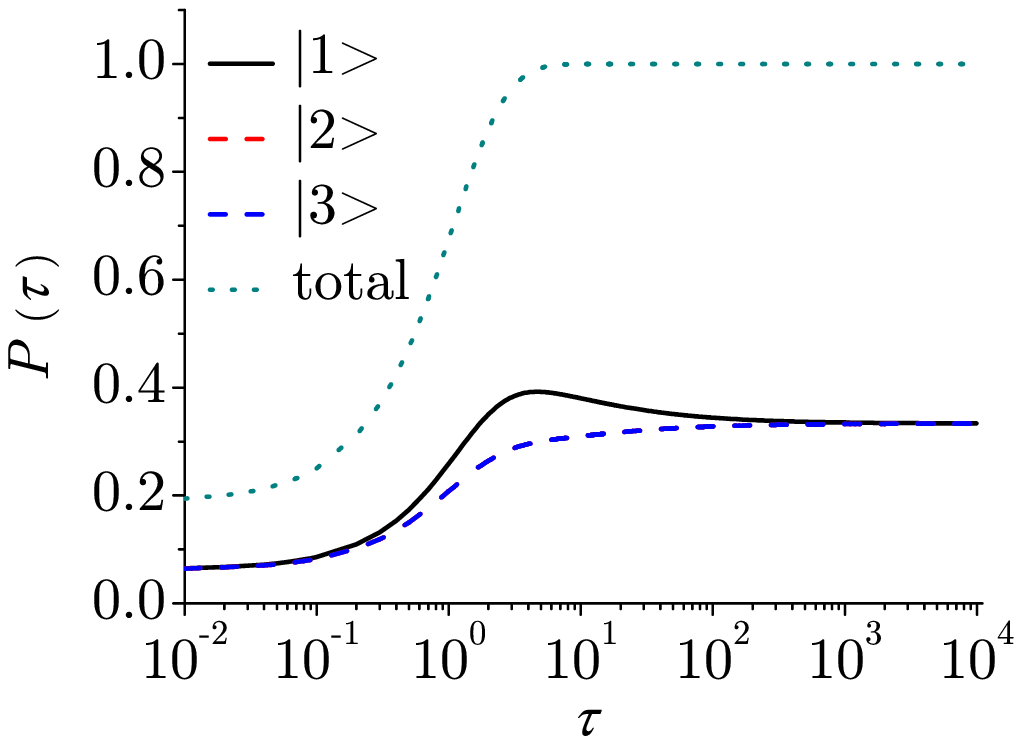}
\end{center}
\end{minipage}
\caption{
Annealing-time dependence of the final probability that the system is
in any one of the ground states. Left panel: Data for the five-spin
model using QA. Only the states $|2\rangle$ and $|3\rangle$ (and
their reversals $|\bar{2}\rangle$ and $|\bar{3}\rangle$) are reached
for large $\tau$.  Right panel: SA finds all the states
with equal probability.
\label{fig:5spin-P}}
\end{figure}

The situation changes significantly for the present degenerate case
as illustrated in figure \ref{fig:e-spectrum}, which depicts the
instantaneous energy spectrum.
Some of the excited states reach the final ground state as $t/\tau\to
1$.  In particular, the instantaneous ground state configurations
have been found to be continuously connected to a special symmetric
combination of four of the final ground states at $t=\tau$,
$|2\rangle+|\bar{2}\rangle+|3\rangle+|\bar{3}\rangle$, whereas the
other two states $|1\rangle$ and $|\bar{1}\rangle$ are out of reach as
long as the system faithfully follows the instantaneous ground state
($\tau\gg 1$). A relatively quick time evolution with an intermediate
value of $\tau$ may catch the missed ground states.  However, there is
no guarantee that the obtained state using this procedure is a true
ground state since some of the final excited states may be reached.
As shown in the left panel of figure \ref{fig:5spin-P}, intermediate
values of $\tau$ around 10 give almost an even probability to all
the true ground states, an ideal situation.  However, the problem is
that we do not know such an appropriate value of $\tau$ beforehand.
In contrast, the right panel of figure \ref{fig:5spin-P} shows
the result of SA by a direct numerical integration of the master
equation, in which all the states are reached evenly in the limit of
large $\tau$.
Figure \ref{fig:e-spectrum} suggests that it might be plausible to
start from one of the low-lying excited states of $H_1$ to reach the
missed ground state.  However, such a process has also been found to
cause similar problems as above. We therefore conclude that QA is not
suitable to find all degenerate ground-state configurations of the
target system $H_0$, at least in the present example. This aspect is
to be contrasted with SA, in which infinitely-slow annealing of the
temperature certainly finds all ground states with equal probability
as assured by the theorem of Geman and Geman \cite{geman:84}.

\begin{figure}[htb]
\begin{minipage}{0.50\hsize}
\begin{center}
\includegraphics[width=1\linewidth]{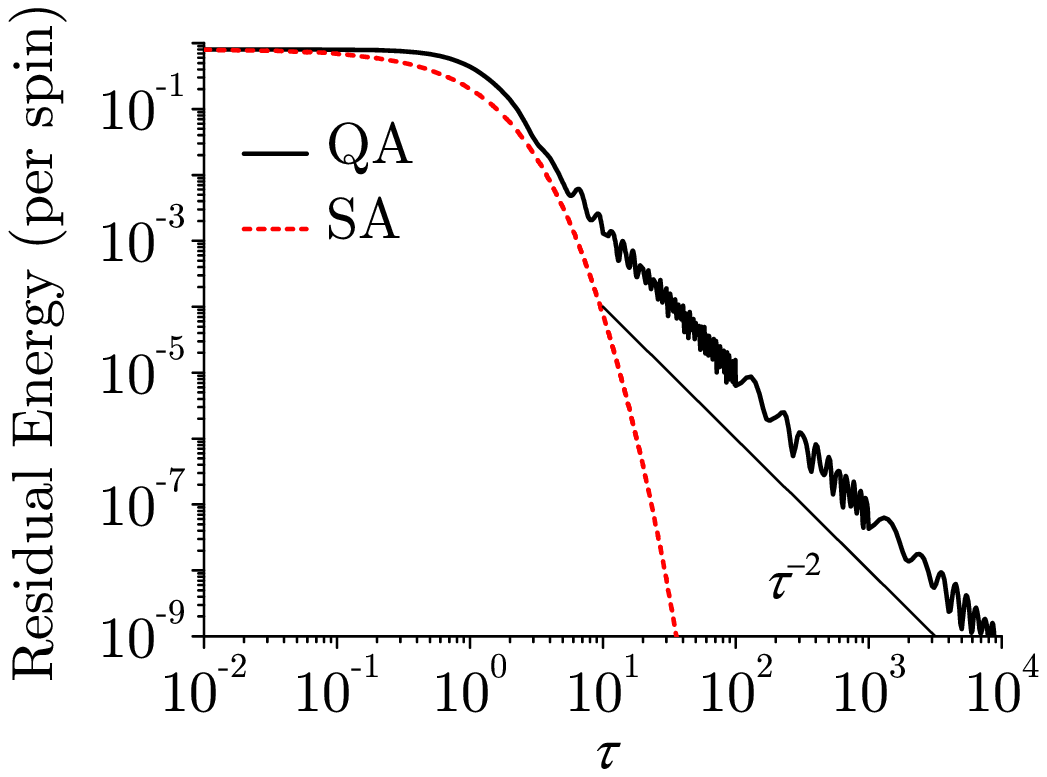}
\end{center}
\end{minipage}
\begin{minipage}{0.50\hsize}
\begin{center}
\includegraphics[width=1\linewidth]{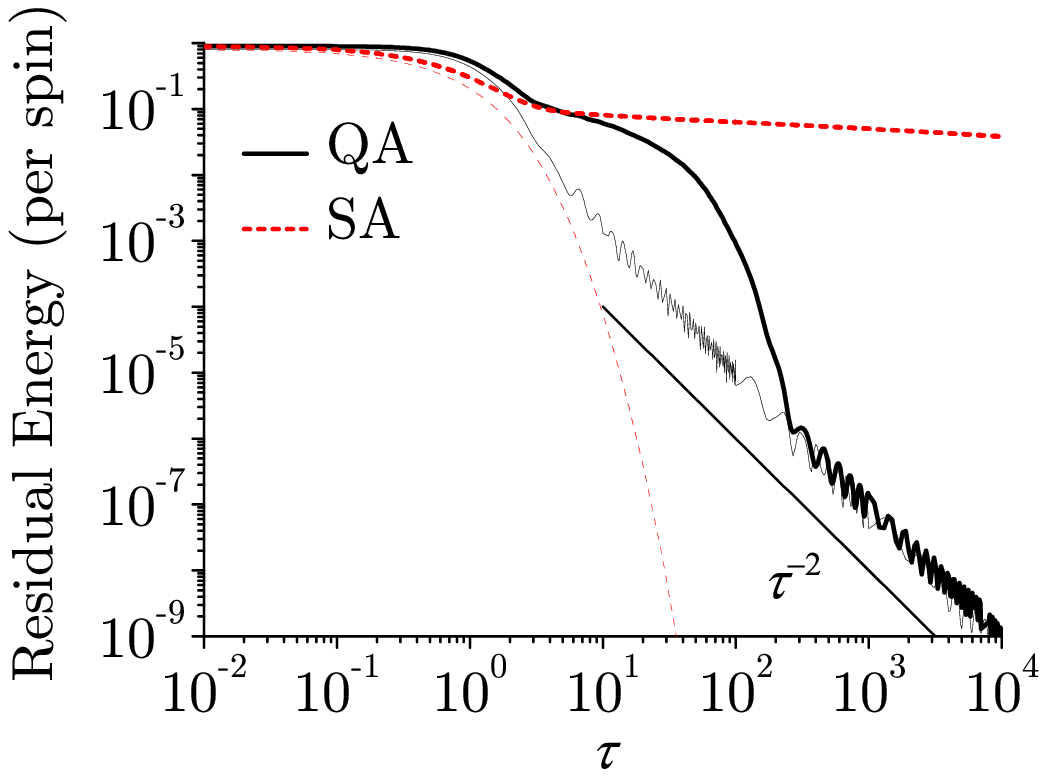}
\end{center}
\end{minipage}
\caption{
Residual energy per spin as a function of the annealing time $\tau$ for
the five-spin toy model. Left panel: Degenerate case. Right panel: A
small perturbation $h = 0.10$ [see equation (\ref{longitudinal-field})]
has been added to lift the overall spin-reversal symmetry and thus
break the degeneracy. While QA is rather robust against the inclusion
of a longitudinal field term and the residual energy per spin decays
in both cases $\sim \tau^{-2}$ for large $\tau$, SA seems not to
converge after the inclusion of a field term. The thin lines on the
right panel represent the results in zero field (from the left panel)
for comparison.  \label{fig:5-spin-res-energy}}
\end{figure}

QA nevertheless shows astounding robustness against a
small perturbation that lifts part of the degeneracy if our
interest is in the value of the ground-state energy.  Figure
\ref{fig:5-spin-res-energy} depicts the residual energy---the
difference between the obtained approximate energy and the true
ground-state energy---as a function of the annealing time $\tau$.

Data using SA are also shown in figure \ref{fig:5-spin-res-energy}
using an annealing schedule of temperature  $T=(\tau -t)/t$,
corresponding to the ratio of the first and the second terms on
the right-hand side of equation (\ref{full-Hamiltonian}). In the
degenerate case (left panel) SA outperforms QA since the residual
energy decays more rapidly using SA. However, if we apply a small
longitudinal field to $H_0$
\begin{equation}
  H_2=-h\sum_{i=1}^5 \sigma_i^z
 \label{longitudinal-field}
\end{equation}
and regard $H_0+H_2$ as the target Hamiltonian to be minimized,
the situation changes drastically: The convergence of SA slows down
significantly while the convergence of QA remains almost intact (right
panel of figure  \ref{fig:5-spin-res-energy}).  As already observed in
the simple double-well potential problem \cite{stella:05}, the energy
barrier between the two almost degenerate states of $H_0+H_2$ may be
too high to be surmountable by SA whereas the width of the barrier
remains thin enough to allow for quantum tunnelling to keep QA working.

\begin{figure}[htb]
\begin{minipage}{0.50\hsize}
\begin{center}
\includegraphics[width=1\linewidth]{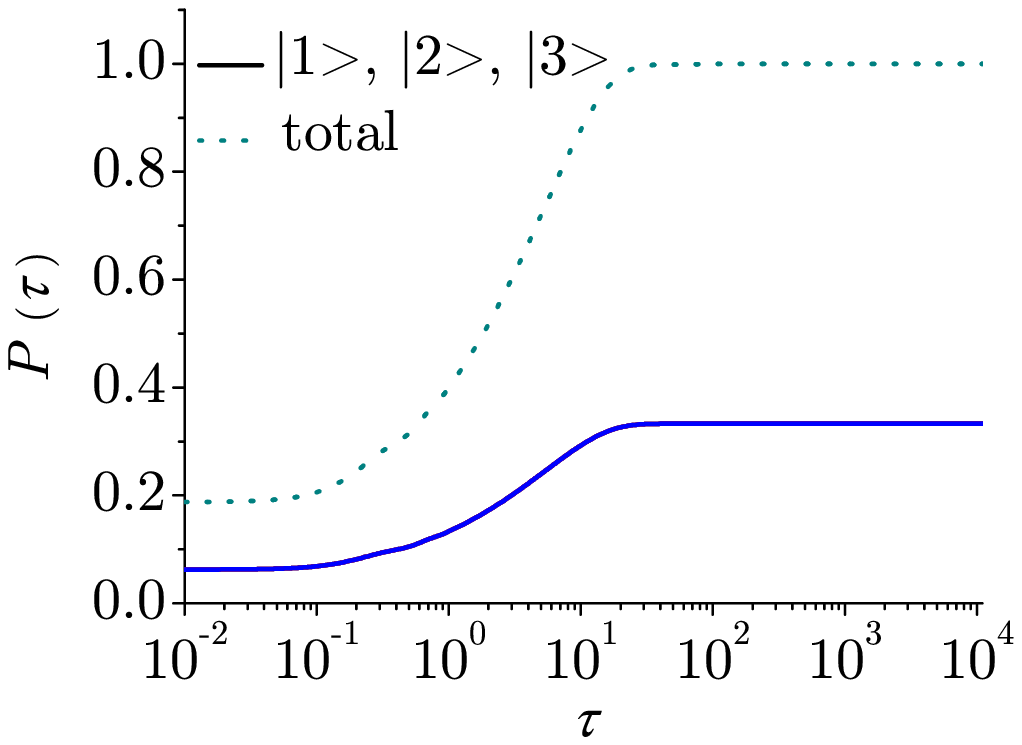}
\end{center}
\end{minipage}
\begin{minipage}{0.50\hsize}
\begin{center}
\includegraphics[width=1\linewidth]{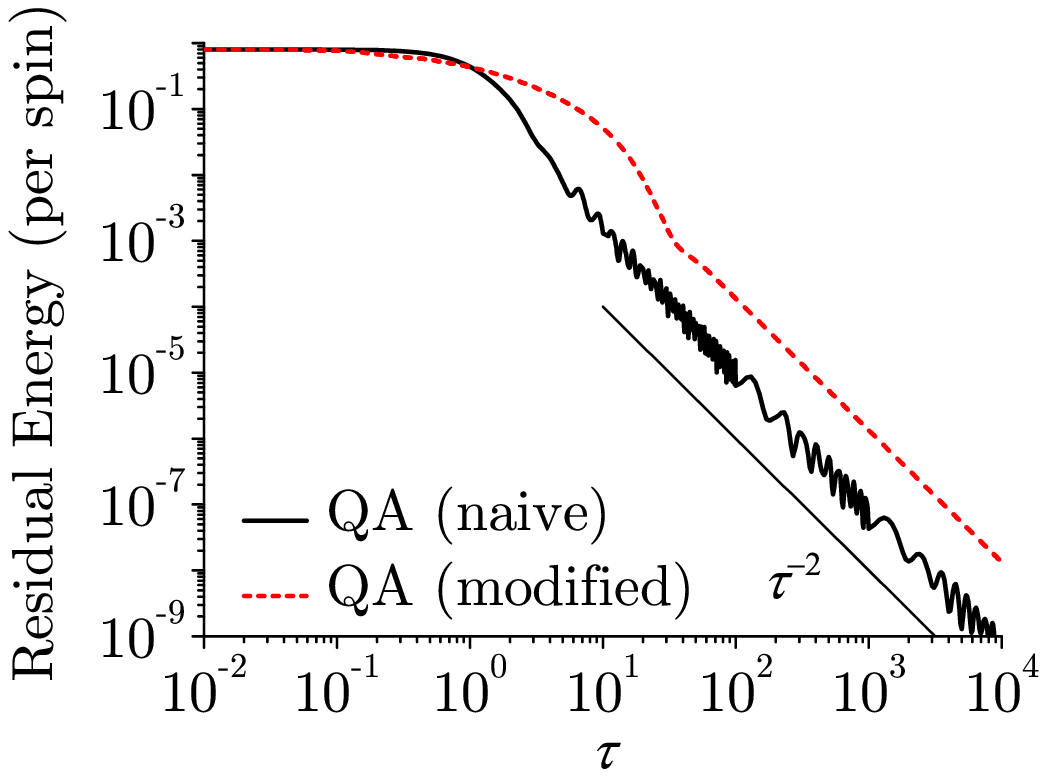}
\end{center}
\end{minipage}
\caption{\label{fig:5-spin-mod}
Left panel: Annealing-time dependence of the final probability using
the modified Hamiltonian which allows for transitions between states.
The data for $\left| 1 \right\rangle$, $\left| 2 \right\rangle$
and $\left| 3 \right\rangle$ fully overlap. Right panel: Residual
energy per spin as a function of the annealing time $\tau$ for the
modified Hamiltonian. The solid line corresponds to QA using only
a transverse field and is depicted for comparison (data from figure
\ref{fig:5-spin-res-energy}).
}
\end{figure}

It is relatively straightforward to modify the type of quantum
fluctuations to force the system to finally reach all ground states
with equal probability. The following term
\begin{equation}
H_1 ^\prime    =  - \sum\limits_i {\sigma _i^x }  - \sum\limits_{\left\langle {i,j} \right\rangle } {\sigma _i^x \sigma _j^x }  - \sum\limits_{\left\langle {i,j,k} \right\rangle } {\sigma _i^x \sigma _j^x \sigma _k^x }  -  \cdots
\label{H1p}
\end{equation}

is used instead of $H_1$ of equation (\ref{H1}) to induce direct
quantum transitions between all states. This modified quantum term is
not unnatural in view of the quantum-annealing version of the Grover
algorithm \cite{roland:02}, in which the quantum term connects all
states with equal transition amplitude. The result for the five-spin
model is shown in figure \ref{fig:5-spin-mod} (left panel), in which
all states are found with equal probability as expected. As depicted
in the right panel of figure \ref{fig:5-spin-mod}, the scaling of
the residual energy still shows the $1/\tau^2$ behavior, although an
order of magnitude slower. A drawback of this method is that it is not
easily implemented by Monte Carlo simulations because of complicated
interactions appearing in the path-integral representation of equation
(\ref{H1p}). We therefore restrict the use of this method to the
present section, and Monte Carlo simulations in the next section will
be carried out only for the simple transverse-field case of equation
(\ref{H1}).

\section{Monte Carlo simulations for larger systems}
\label{sec:large}

\begin{figure}[h]
\scalebox{0.18}{\input{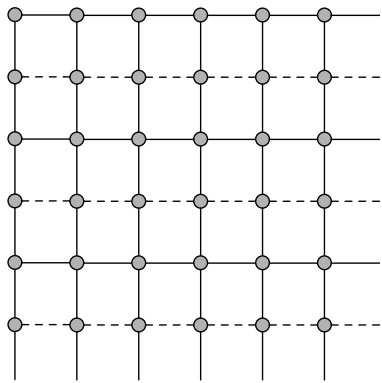}}
\begin{minipage}[b]{25pc}\caption{\label{fig:ff-lattice}
Fully-frustrated Ising model with $N = 36$ spins (dots) and periodic
boundary conditions. The system has $45088$ degenerate ground states
excluding spin-reversal symmetry. For comparison, for $L=4$ the
system has only $136$ degenerate ground states.  The horizontal bonds
alternate between ferromagnetic (full lines) and antiferromagnetic
(dashed lines) signs. The vertical bonds are all ferromagnetic. This
ensures that the product of all bonds around any plaquette is negative,
i.e., the system is maximally frustrated.
}
\end{minipage}
\end{figure}

The simplest Ising model which possesses an exponentially-large
ground-state degeneracy (in the system size) is the two-dimensional
Villain fully-frustrated Ising model \cite{villain:77}. The Ising
model is defined on a square lattice of size $N = L\times L$ and has
alternating ferromagnetic and antiferromagnetic interactions in the
horizontal direction and ferromagnetic interactions in the vertical
direction; see figure \ref{fig:ff-lattice}.  We use periodic boundary
conditions and show the data for $L = 4$ and $6$ with $M = 1/T = 100$,
where $M$ is the number of Trotter slices and $T$ the temperature.
We follow Ref.~\cite{kadowaki:98a} and choose $T=1/M$ to ensure
an optimal performance of quantum Monte Carlo simulations applied
to quantum annealing.  A transverse field $H_1 = -\Gamma \left(
t \right) \sum_{i} \sigma_i^x$ is applied, followed by a pre-annealing
of the system by decreasing the temperature as $T=100/Mt$ from $t=0$
to $100$.  After the pre-annealing, the transverse field is decreased ($t$
is set to $0$) by the schedule $\Gamma \left( t \right) =(\tau-t)/t$
from $t=0$ to $\tau$ with $T$ fixed.  The ground-state energy of the
target Hamiltonian $H_0$ is given by $-L^2$. We can therefore check
easily if each Trotter slice has reached one of the ground states.

\begin{figure}[htb]
\begin{minipage}{0.515\hsize}
\begin{flushleft}
\includegraphics[width=1\linewidth]{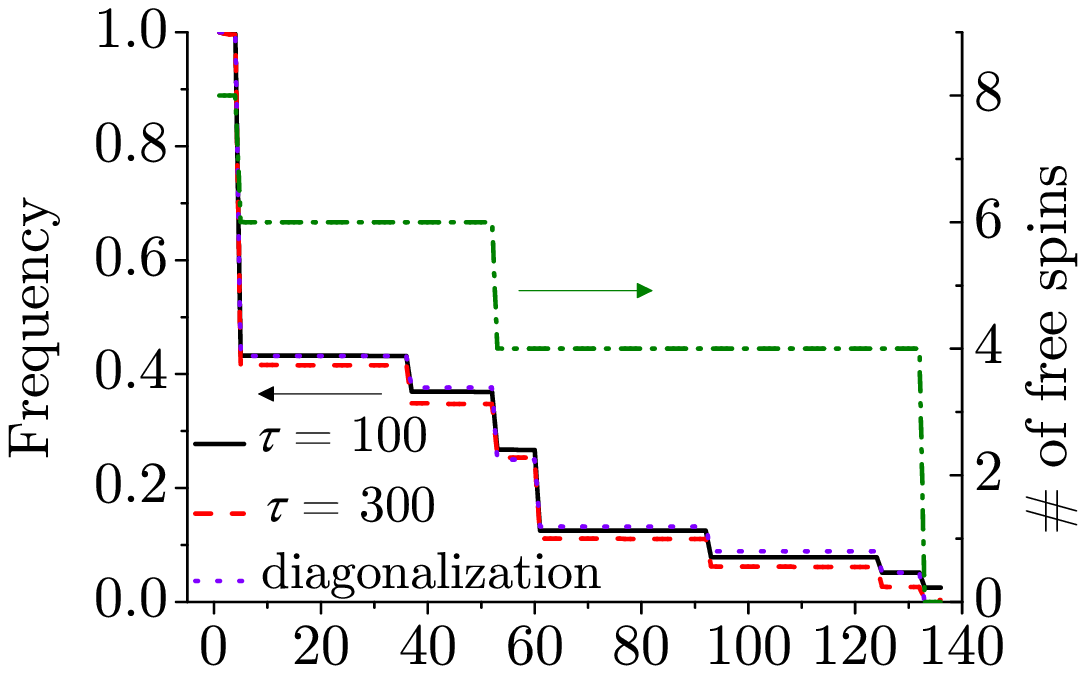}
\end{flushleft}
\end{minipage}
\begin{minipage}{0.485\hsize}
\begin{flushleft}
\includegraphics[width=1\linewidth]{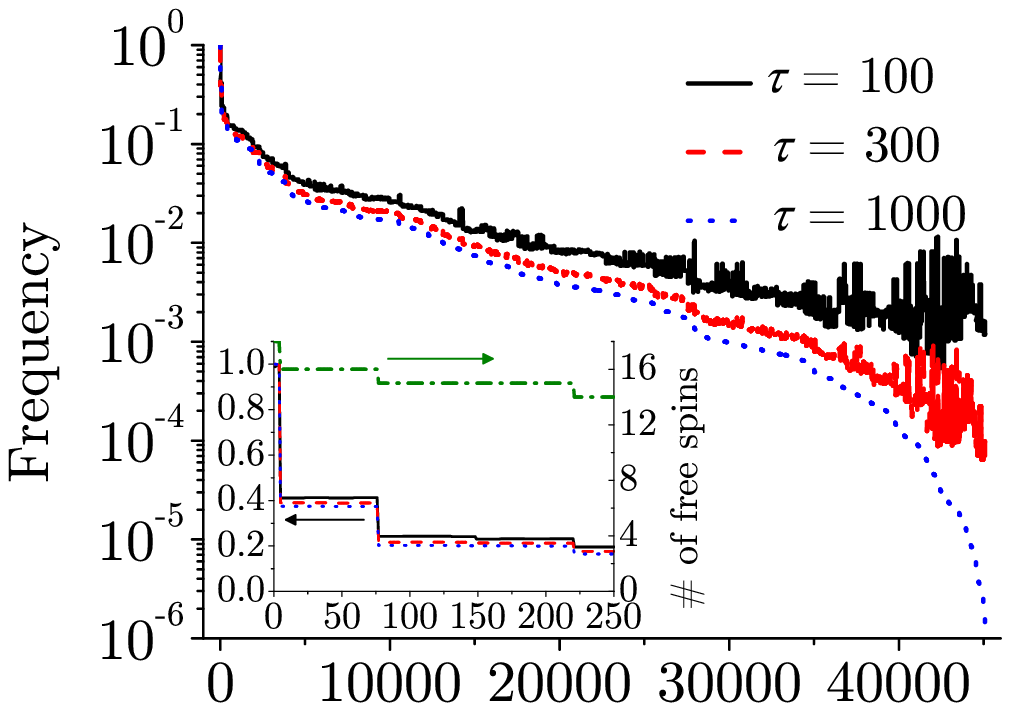}
\end{flushleft}
\end{minipage}
\caption{
Histograms of the relative frequency that a given ground state is
reached by QA. In the abscissa the ground states corresponding to all
annealing times are numbered according to the relative frequency of
$\tau = 300$ for $L=4$ (left panel) and $\tau = 1000$ for $L = 6$
(right panel) to be reached.  The chained lines denote the number
of free spins (the spins that can be flipped without energy cost)
corresponding to the abscissa. The inset of the right panel is for
the high-frequency configurations of the system size $L = 6$.
\label{fig:ff-qahits}}
\end{figure}

\begin{figure}[htb]
\begin{minipage}{0.50\hsize}
\begin{flushleft}
\includegraphics[width=1\linewidth]{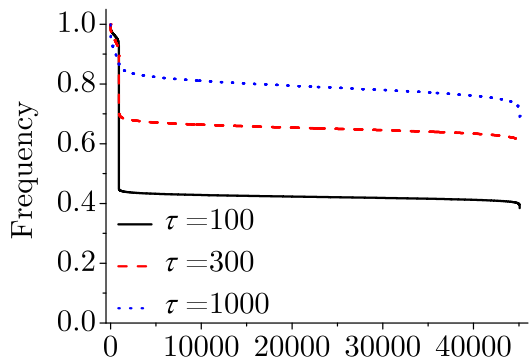}
\end{flushleft}
\end{minipage}
\begin{minipage}{0.50\hsize}
\caption{
Histograms of the relative frequency by SA. Most ground states are
found evenly and all ground states can be reached at least 40\%
of the time.
\label{fig:ff-sahits}}
\end{minipage}
\end{figure}

Figures \ref{fig:ff-qahits} and \ref{fig:ff-sahits} show 
the relative number of ground-state hits versus
the ground-state numbering (sorted by the number of hits). While
a small part of the total set of ground states are reached very
frequently, some ground states seem exponentially suppressed in
QA. The result is practically independent of the chosen value of the
annealing time $\tau$ in the Monte Carlo simulation and coincides
with data from the diagonalization of the Hamiltonian corresponding to
the adiabatic limit ($\tau \gg 1$) of the Schr\"odinger dynamics. We
may thus conclude that our Monte Carlo simulations with small but
finite $T = 1/M$ approximate well the final state of the adiabatic
Schr\"odinger equation. In figure \ref{fig:ff-sahits} we show the
relative number of ground-state hits versus the ground-state numbering
for SA. In contrast to QA, SA finds almost all ground states with
uniform probability. While some ground states seem to be preferred,
all ground states can be reached with a frequency of at least 40\%
(see also Ref.~\cite{moreno:03}).

\begin{figure}[tb]
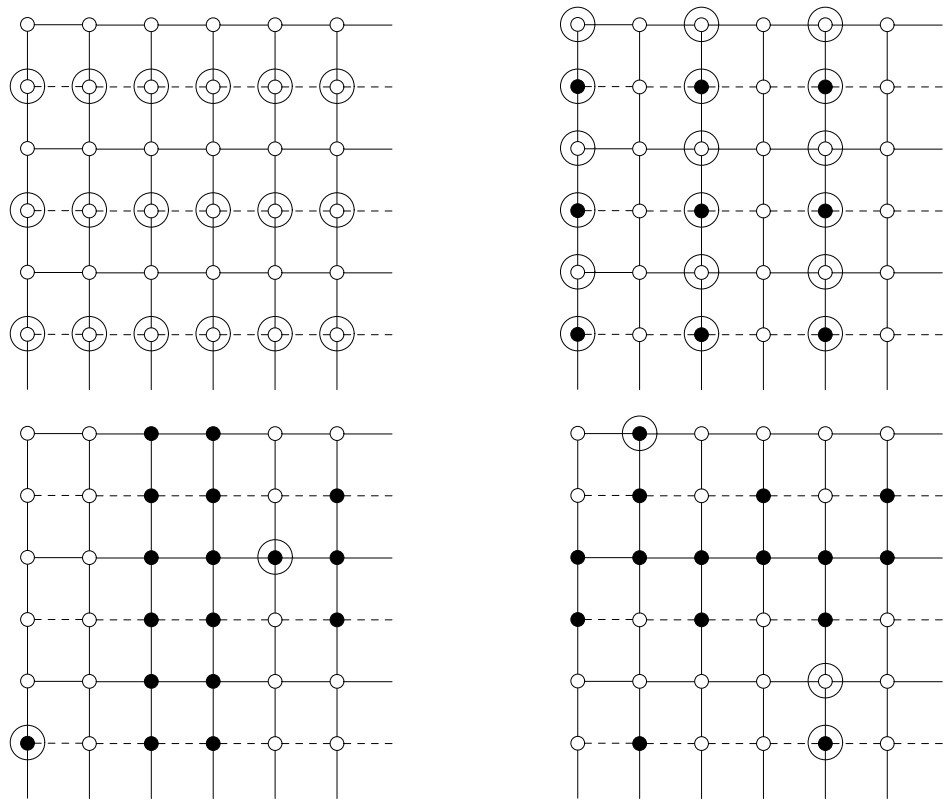

\begin{center}
\scalebox{0.18}{\input{ff_high.pic}}
\end{center}
\begin{center}
\scalebox{0.18}{\input{ff_low.pic}}
\end{center}
\caption{\label{fig:ff-examples}
Typical examples of the high-frequency configurations (top panels)
and low-frequency configurations (bottom panels). Filled and open
circles denote up and down spins, respectively. Circled spins can be
flipped independently without energy changes (free spins).
}
\end{figure}

It is interesting to study the microscopic configurations of those
states with high frequency. Figure \ref{fig:ff-examples} shows typical
spin configurations of states with high frequency (top panel) and low
frequency (bottom panel). The number of free spins (the spins that
can be flipped without energy cost) is much larger in the former
that in the latter. In figure \ref{fig:ff-qahits} we also show the
number of free spins which clearly shows that the states with more
free spins are reached more frequently.  A classical ground state
with a larger number of free spins would have a lower energy than
a state with fewer free spins under a small perturbation of the
transverse field because the spins flipped by the transverse field
are less likely to be free spins in the latter case with fewer free
spins, resulting in high energies. An alternative view is provided
if we map a ground-state spin configuration onto a (directed) dimer
configuration on the dual lattice by connecting neighbouring plaquettes
across an unsatisfied bond \cite{moessner:01,moessner:03}.  The flip
of a free spin in real space corresponds to a dimer plaquette flip.
Then the high-frequency configurations in the top panel of figure
\ref{fig:ff-examples} correspond to dimer configurations with low
winding numbers and the low-frequency configurations to large winding
numbers \footnote{We thank one of the referees for pointing this out.};
the former being easier to flip.

We should, however, be careful to assume that these pictures always
explains the situation because we have found a counterexample in
the $\pm J$ Ising model with $L=10$.  A possible reason is that the
number of free spins is relatively small for this spin-glass system
in comparison with the Villain model, which would lead to weaker
effects of spin flips due to quantum fluctuations.  Further studies
are required to clarify this point.

\begin{figure}[tb]
\begin{minipage}{0.50\hsize}
\begin{flushleft}
\begin{overpic}[trim = 0 20 0 0, width=1\linewidth]{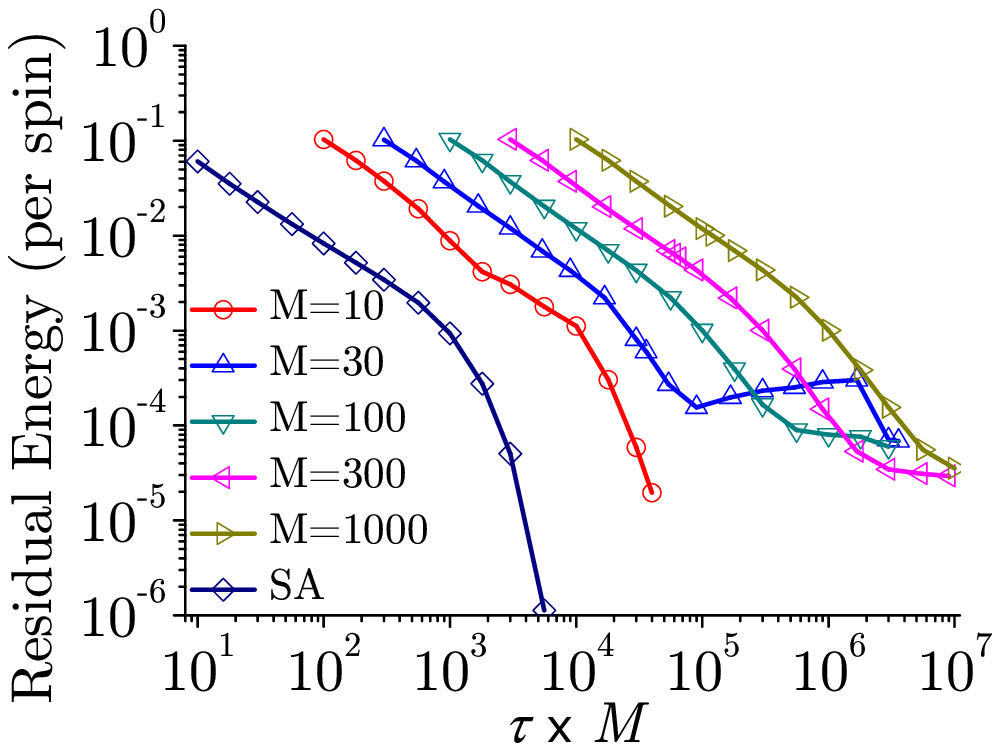}
\put(75, 60){Villain}\put(71, 55){(average)}
\end{overpic}
\end{flushleft}
\end{minipage}
\begin{minipage}{0.50\hsize}
\begin{flushleft}
\begin{overpic}[trim = 0 20 0 0, width=1\linewidth]{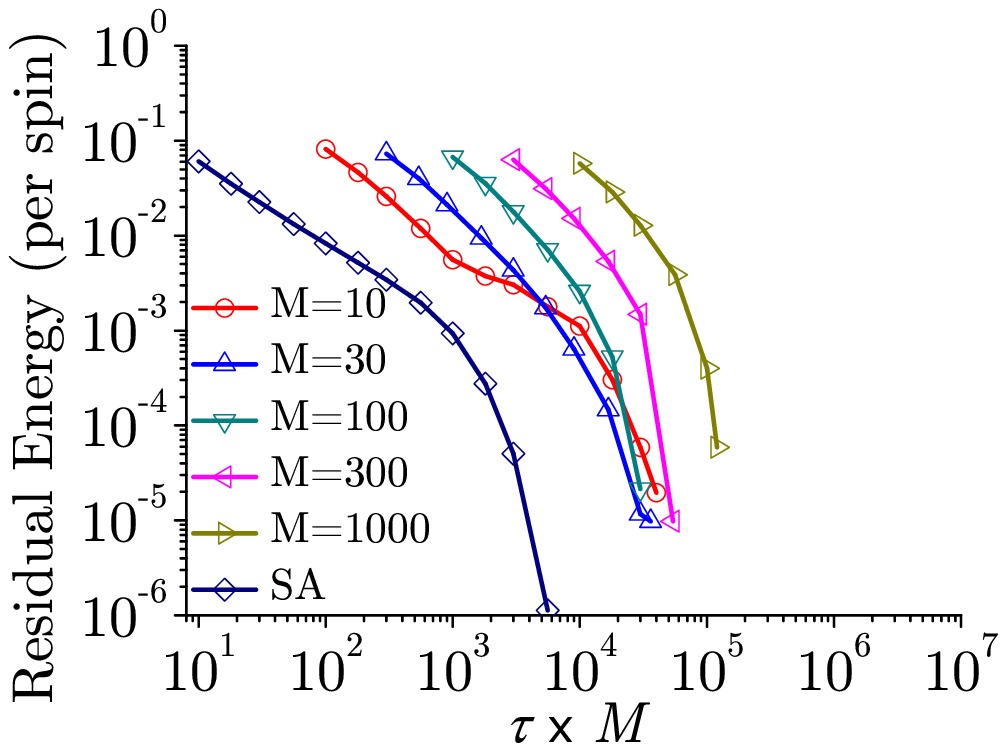}
\put(75, 60){Villain}\put(79, 55){(best)}
\end{overpic}
\end{flushleft}
\end{minipage}
\begin{minipage}{0.50\hsize}
\begin{flushleft}
\begin{overpic}[trim = 0 20 0 0, width=1\linewidth]{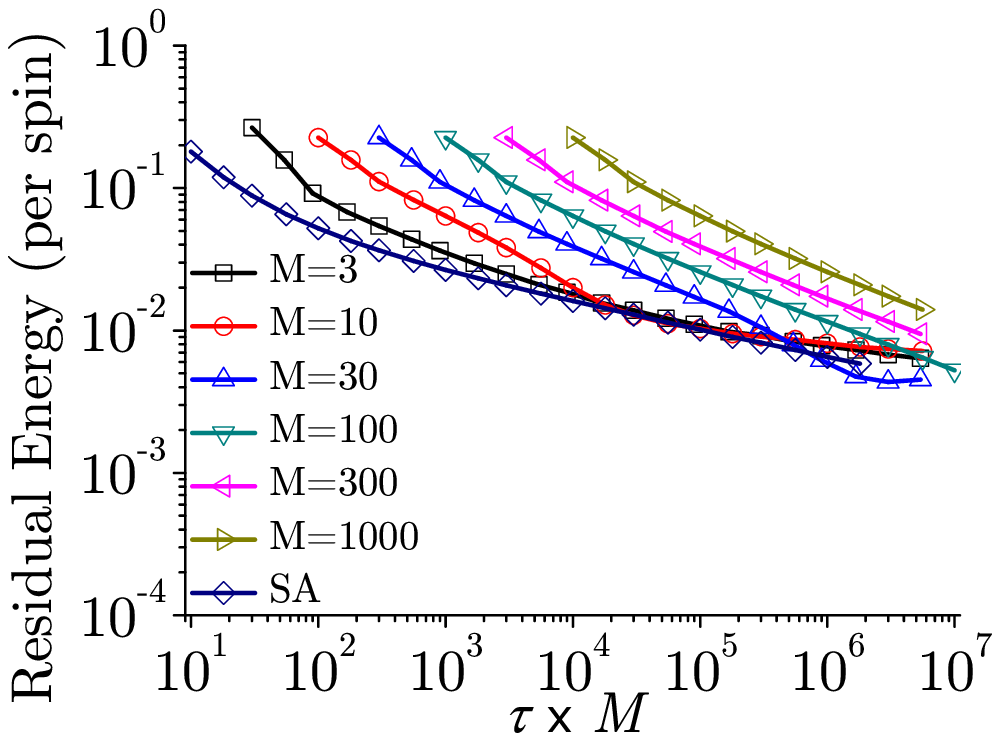}
\put(82, 60){$\pm J$}\put(71, 55){(average)}
\end{overpic}
\end{flushleft}
\end{minipage}
\begin{minipage}{0.50\hsize}
\begin{flushleft}
\begin{overpic}[trim = 0 20 0 0, width=1\linewidth]{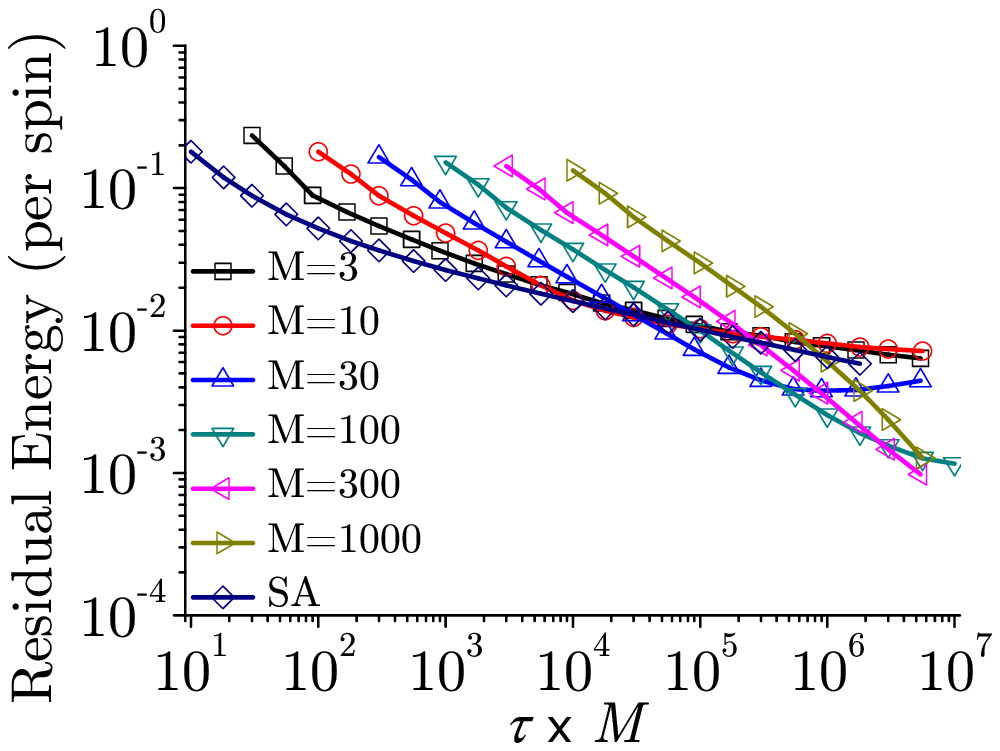}
\put(82, 60){$\pm J$}\put(79, 55){(best)}
\end{overpic}
\end{flushleft}
\end{minipage}
\begin{minipage}{0.50\hsize}
\begin{flushleft}
\begin{overpic}[trim = 0 20 0 0, width=1\linewidth]{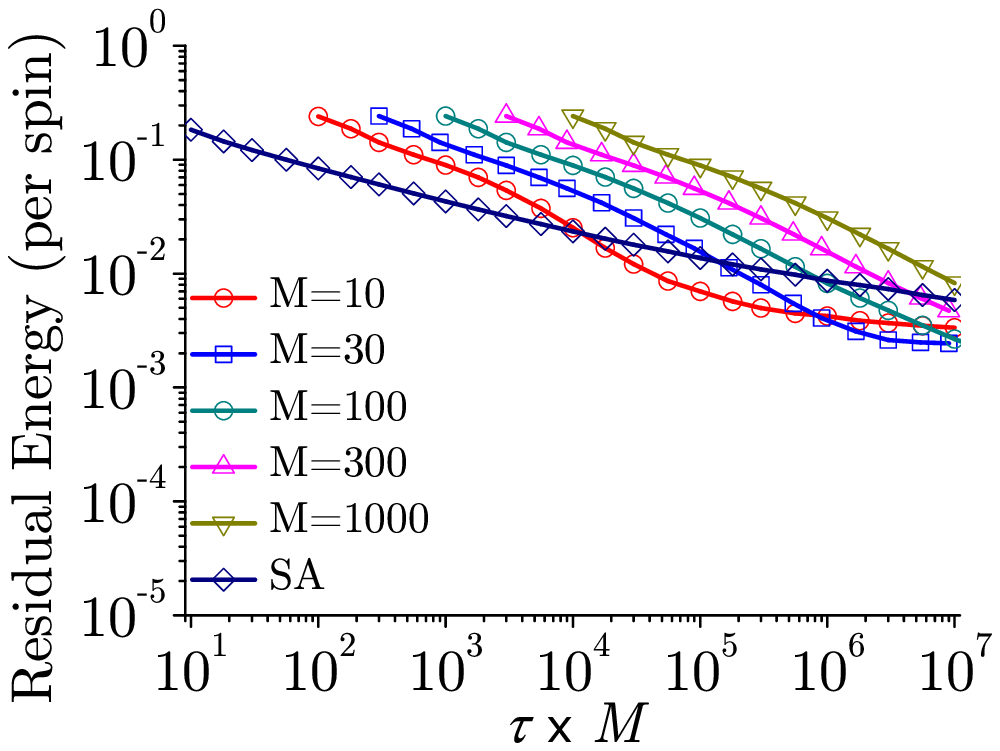}
\put(62, 60){Gaussian SG}\put(71, 55){(average)}
\end{overpic}
\end{flushleft}
\end{minipage}
\begin{minipage}{0.50\hsize}
\begin{flushleft}
\begin{overpic}[trim = 0 20 0 0, width=1\linewidth]{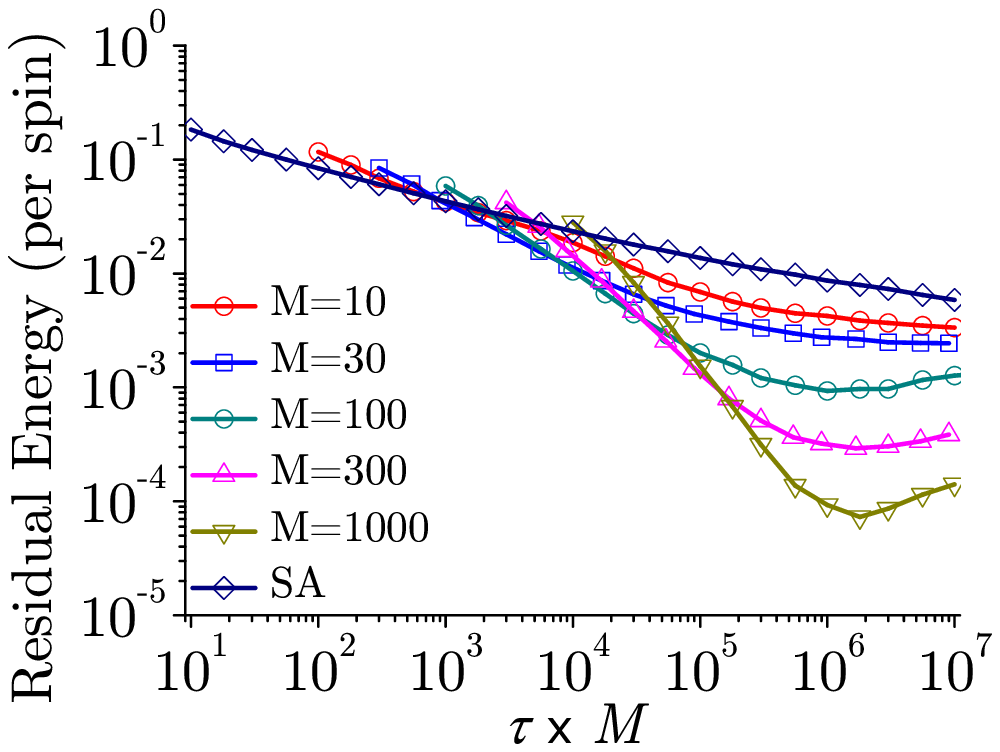}
\put(62, 60){Gaussian SG}\put(79, 55){(best)}
\end{overpic}
\end{flushleft}
\end{minipage}
\caption{\label{fig:resqmc}
Residual energy per spin using QA and SA for the Villain model with
$L=40$ (top panels), the $\pm J$ model with $L=40$ (middle panels)
and the Gaussian spin glass with $L=10$ (bottom panels).  The number
of samples for the bond randomness is $50$ for the $\pm J $ model and
$100$ for the Gaussian model, and the number of runs are over $100$
for the random seedings. We take averages over all samples and
calculate the average energy (left panels) and lowest energy
(right panels) among all Trotter slices. Errors are smaller than the symbols.
}
\end{figure}

We next analyze the residual energy. The top two panels of figure
\ref{fig:resqmc} show the results for the Villain model with the system
size $L=40$ for QA and SA. The abscissa is the annealing time $\tau$
(Monte Carlo steps) multiplied by the number of Trotter slices $M$
for a fair comparison between QA and SA. The left panel is for the average
final energy over all Trotter slices and the right panel represents
the lowest energy among all Trotter slices.  The data for SA follow
a rapid decrease beyond $\tau \approx 2\times 10^3$, whereas QA (the
average energy) stays unimproved beyond this region.
While QA for the best energy keeps improving further at lower $T$ (larger $M$),
QA does not surpass SA in this annealing time region.

Finally, we also show the data for the two-dimensional $\pm J$
Ising spin glass (middle panels) and Gaussian spin glass with a
weak longitudinal field ($h = 0.1$) (bottom panels) in order to  
remove a trivial degeneracy. In fact, the latter model does not 
have a ground-state degeneracy whereas the Villain model and the 
$\pm J$ model have quite a large number of ground
states. In the $\pm J$ and the Gaussian cases the situation is quite
different, since the residual energy using QA decreases more rapidly
than SA. Furthermore, the performance of QA improves for smaller
$T$ (larger $M$) in the large-$\tau$ region. At high temperature
annealing (smaller $M$) it seems that the data of QA are saturated
for large $\tau$. This saturation may reflect the finiteness of
the temperature of QA. For the $\pm J$ and Gaussian models it
seems that QA is a powerful tool, as witnessed in previous studies
\cite{kadowaki:98,das:05,santoro:02,martonak:02,santoro:06,das:08,
kadowaki:98a}.

\section{Conclusion}
\label{sec:concl}

We have studied the performance of QA for systems with degeneracy
in the ground state of the target Hamiltonian.  Our results show
that naive QA with quantum effects induced by a transverse field
reaches only a limited part of the set of degenerate ground-state
configurations and misses the other states.  The instantaneous energy
spectrum as a function of time, figure \ref{fig:e-spectrum}, is a
useful tool to understand the situation: some of the excited states
merge into the final ground states as $t/\tau\to 1$.  It is usually
difficult, however, to initially select the appropriate excited states
which reach the ground state when $t/\tau\to 1$.  We therefore have to
be biased a priori in the state search by QA for degenerate cases. An
introduction of quantum transitions to all other states significantly
improves the situation, leading to a uniform probability to find
degenerate ground states, but with the penalty of slower convergence.
Simulations of the two-dimensional Villain model show that certain
ground-state configurations are exponentially suppressed when using
QA, whereas this is not the case for SA.  It has been found that the
number of free spins is closely related to the frequency that the
ground states are found by QA. A ground state with a larger number of
free spins is more stable than the others.  On the other hand, if the
problem is to only determine the ground-state energy, QA is often (but
not necessarily always) an efficient method, as exemplified in figures
\ref{fig:5-spin-res-energy} and \ref{fig:resqmc} where the residual
energies are shown. Thus, it is necessary to construct an analytical
theory---based on the numerical evidence---to establish criteria when to
(and when not to) use QA.

\ack

We thank Dr.~Shu Tanaka for bringing the effects of free spins
to our attention (section \ref{sec:large}).  We would also like
to thank Prof.~Roderich Moessner and Matthias Troyer for useful
discussions. The present work was supported by the Grant-in-Aid for
Scientific Research for the priority area `Deepening and expansion
of statistical-mechanical informatics' (DEX-SMI) and CREST, JST.
H.G.K.~acknowledges support from the Swiss National Science Foundation
under Grant No.~PP002-114713. Part of the simulations were performed
on the ETH Z\"urich brutus cluster.

\section*{References}

\bibliography{refs}

\end{document}